\documentclass[11pt,fleqn]{article}
\usepackage{amssymb}

\usepackage{amsmath}

\usepackage{amsmath,amssymb,amsthm}

\begin{document}
{\Large
\begin{center}
{\Large \textbf{Conditional Symmetry and Reductions for the
Two-Dimensional Nonlinear Wave Equation. I. General Case.}}

\vskip 20pt {\large \textbf{Irina YEHORCHENKO}}

\vskip 20pt {Institute of Mathematics of NAS Ukraine, 3
Tereshchenkivs'ka Str., 01601 Kyiv-4, Ukraine} \\
E-mail: iyegorch@imath.kiev.ua
\end{center}

\vskip 50pt
\begin{abstract}
We present classification of $Q$-conditional symmetries for
the two-dimensional nonlinear wave equations
$u_{tt} - u_{xx} = F(t,x,u)$
and the reductions corresponding to these nonlinear symmetries.
Classification of inequivalent reductions is discussed.
\end{abstract}

\section{Introduction}
Following \cite{CS NLWE:IYpreprint09},
we discuss conditional symmetries and reductions
of the two-dimensional nonlinear
wave equation
\begin{equation} \label{CS NLWE:Fwave}
u_{tt} - u_{xx} = F(t,x,u)
\end{equation}
for the real-valued function $u = u(t,x)$; $t$ is the time
variable, $x$ is the space variable. In the equation above and further we will use the
following designations for the partial derivatives:
\[
u_t = - \frac{\partial u}{\partial t}; \\ u_x = - \frac{\partial
u}{\partial x}; \\ u_{tt}=\frac{\partial^2 u}{\partial t^2}; \\
u_{xt}=u_{tx}=\frac{\partial^2 u}{\partial t \partial x}; \\
u_{xx}=\frac{\partial^2 u}{\partial x^2}.
\]

Note that the general equation in the class (\ref{CS NLWE:Fwave}) has no invariance operators;
however, many well-known particular
cases have wide symmetry algebras, see e.g. \cite{CS
NLWE:FSerov-dA}.

The maximal invariance algebra of the equation (\ref{CS
NLWE:Fwave}) with general $F=F(u)$ is the Poincar\'e algebra $AP(1,1)$ with the basis
operators
\begin{gather*}
 p_t  =  \frac {\partial}{\partial t}, \quad p_x  =  \frac {\partial}{\partial x},
 \quad J = t p_x + x p_t.
\end{gather*}

The invariance algebras of the equation (\ref{CS NLWE:Fwave}) will
also include dilation operators e.g. for $F=\lambda u^k$ or $F=\lambda
\exp u$. Equations (\ref{CS NLWE:Fwave}) with e.g.
$F=0$ and $F=\lambda \exp u$ have infinite-dimensional symmetry algebras.

Similarity solutions for the equation (\ref{CS NLWE:Fwave}) can
be found by symmetry reduction with respect to non-equivalent
subalgebras of its invariance algebras. For studies of symmetry and non-classical solutions of the nonlinear
wave equation for various space dimensions see \cite{CS NLWE:FSerov-dA}--\cite{CS NLWE:Bizon}.

Here we present results on classification of $Q$-conditional
symmetries for the equation (\ref{CS NLWE:Fwave}) and the relevant reductions
in the meaningful cases.

It seems that investigation of conditional symmetry now has fallen out of the mainstream
of the symmetry analysis of PDE. We would guess that the reason is that practically
all interesting equations for which the problem is manageable (mostly for the evolution equations)
have been studied already. However, we believe that this problem remains relevant - first, with respect to
investigation of "difficult, but interesting" equations
(e.g. equations with highest derivatives for all variables of the same order, such as the nonlinear
wave equation under study),
and with respect to investigation of various related aspects
(e.g. geometrical aspects and equivalence).

\section{What we mean by conditional symmetry}

Conditional symmetry in general (additional
invariance under arbitrary additional condition) and a narrower
concept of the  $Q$-conditional invariance (the additional
condition has the form $Qu=0$) were initially discussed in the
papers \cite{CS NLWE:OlverRosenau}-\cite{CS NLWE:LeviWinternitz}.
Later numerous authors developed these ideas into theory and
a number of algorithms for studying symmetry properties of
equations of mathematical physics. The importance of investigation
of the  $Q$-conditional symmetry stems from equivalence of the
$Q$-conditional invariance and reducibility of the equations by
means of ansatzes determined by such operators $Q$ (see \cite{CS
NLWE:zhdanov&tsyfra&popovych99}).

Here we will use the following definition of the $Q$-conditional
symmetry:

\noindent {\bf Definition 1.} {\it The equation
$\Phi(x,u,\underset{1}{u},\ldots , \underset{l}{u})=0$, where
$\underset{k}{u}$ is the set of all $k$th-order partial
derivatives of the function $u=(u^1,u^2,\ldots ,u^m)$, is called
$Q$-conditionally invariant \cite{CS NLWE:FSS} under the operator
\[
Q=\xi ^i(x,u)\partial_{x_i}+\eta ^r(x,u)\partial_{u^r}\nonumber
\]
if there is an additional condition
\begin{equation}
Qu=0, \label{CS NLWE:G=0}
\end{equation}
such that the system of two equations $\Phi=0$, $Qu=O$ is
invariant under the operator $Q$. All differential consequences
of the condition $Qu=0$ shall be taken into account up to the
order $l-1$.}

This definition of the conditional invariance of some equation implies
in reality a Lie symmetry (see e.g. the classical
texts \cite{{CS NLWE:Ovs-eng},{CS NLWE:Olver1},{CS
NLWE:BlumanKumeiBook}}) of the same equation with a certain
additional condition. Conditional symmetries of the
multidimensional nonlinear wave equations are specifically
discussed in \cite{{CS NLWE:FSerov88},{CS NLWE:BarMosk},{CS
NLWE:zhdanov-panchakBoxu}}.

\section{Previous work on the problem}

The particular problem we discuss here was first mentioned by
P.Clarkson and E. Mansfield in \cite{CS NLWE:Clarkson Mansfield CS
NLWE} (the case $f=f(u)$), where the relevant determining
equations were written out but not solved, and studies were continued by M.
Euler and N. Euler in \cite{CS NLWE:Euler CS NLWE}. In the latter
paper the determining conditions for the $Q$-conditional
invariance were taken without consideration of the differential
consequences of the condition $Qu=0$, so the resulting operators
did not actually present the solution of the problem.

Following \cite{CS NLWE:Clarkson Mansfield CS NLWE}, we will
consider the equation (in conic variables) equivalent
to (\ref{CS NLWE:Fwave}) of the form
\begin{equation} \label{CS NLWE:yz-wave}
u_{yz} = f(y,z,u).
\end{equation}

We search for the operators of $Q$-conditional invariance in the
form
\begin{equation} \label{CS NLWE:Q-operator}
Q=a(y,z,u)\partial_y+b(y,z,u)\partial_z+c(y,z,u)\partial_u.
\end{equation}

\section{Background of Classification of Conditional Symmetries}
We will base our classification on the procedure of solution of the determining
equations for conditional symmetry and then study equivalence within the
resulting classes.

Anyway, the equivalence group of the system of the equation (\ref{CS NLWE:yz-wave})
and the additional condition
\begin{equation} \label{CS NLWE:Q-condition}
a(y,z,u)u_y+b(y,z,u)u_z+c(y,z,u)=0
\end{equation}

\noindent
determined by the operator (\ref{CS NLWE:Q-operator}) is quite narrow,
and the standard classification procedure may not be appropriate.

Let us note that the concept of equivalence of $Q$-conditional symmetries was
introduced by R. Popovych in \cite{CS NLWE:Popovych2000}.

We study the conditional symmetry for the general case $f=f(y,z,u)$. We will not consider the case $f=0$, as
equation (\ref{CS NLWE:yz-wave}) in this case has a general solution, so its conditional symmetries may be not
quite relevant. There are some interesting special partial cases of the equation (\ref{CS NLWE:yz-wave}), first of all
when $f=f(u)$ and $f=r(y,z)u$ that will be considered in a future paper.

Considering the system (\ref{CS NLWE:yz-wave}), (\ref{CS NLWE:Q-condition}),
we can see three inequivalent cases to be studied separately:

1. $a=0$, $b \ne 0$.

Then we can take
\begin{equation} \label{CS NLWE:Q-operator a=0}
Q=\partial_z+L(y,z,u)\partial_u
\end{equation}

The case $a \ne 0$, $b=0$ is equivalent to $a=0$, $b \ne 0$.

In such cases the additional condition
reduces equation (\ref{CS NLWE:yz-wave})
to a pair of the first-order equations.

2. $a \ne 0$, $b \ne 0$.

In this case we can take
\begin{equation} \label{CS NLWE:Q-operator a ne 0}
Q=\partial_y + K(y,z,u)\partial_z+L(y,z,u)\partial_u.
\end{equation}
where $K(y,z,u)\ne 0$.

This case is obviously the most interesting for consideration.
It might seem appropriate to consider separately the case $c=0$, but it is easy
to check that such systems may be equivalent to the general systems within case 2,
so it should not be considered separately.

3. $a=0$, $b=0$

This case is trivial and in the case if the original equation and the additional condition are
compatible, the additional condition just gives a solution for the equation.

For cases 1-2 the additional condition $Qu=0$ will be represented respectively
by the equations
\begin{equation} \label{CS NLWE:Qu a=0}
u_z=L(y,z,u),
\end{equation}
and
\begin{equation} \label{CS NLWE:Qu a ne 0}
u_y+K(y,z,u)u_z=L(y,z,u).
\end{equation}

We can start with considering of determining equations for the case 2 with $K \ne 0$, having
in mind both cases. The determining equations for the conditional symmetry have the form
\begin{equation}
-K_u^2+K_{uu}K=0, \label{CS NLWE:det eq1}
\end{equation}

\begin{equation}
-K L_{uu}+ \frac{K_u K_y}{K}+\frac{K_u^2 L}{K} +K_u(L_u-K_z)-
K_{uy}-L K_{uu}+KK_{zu}=0, \label{CS NLWE:det eq2}
\end{equation}

\begin{gather}
L_{uy}-L_{uz}K+L_{uu}L - \frac{L_u K_y}{K}+\frac{K_y K_z}{K} -
K_{yz}- \nonumber \\
3K_uf-\frac{K_u L}{K}(L_u-K_z)+K_u L_z -K_{zu}L =0, \label{CS
NLWE:det eq3}
\end{gather}

\begin{equation}
-f_y-Kf_z-Lf_u+L_{yz}+L_{uz}L+L_u f -
\frac{K_y}{K}(L_z-f)-K_zf-\frac{K_u L}{K}(L_z-f)=0 \label{CS
NLWE:det eq4}
\end{equation}

Let us note that these determining equations were first found
in \cite{CS NLWE:Clarkson Mansfield CS NLWE}, though, not studied further.

\section{Conditional Symmetry: Main Results}
{\bf Case 1} - $K=0$ . Here we have
equations
\begin{equation} \label{CS NLWE:c1}
u_y=L, u_{yz}=f.
\end{equation}

The determining equations have the form
\begin{equation}
L_{uy}+L_{uu}L =0, \nonumber
\end{equation}

\begin{equation}
-f_y-Lf_u+L_{yz}+L_{uz}L+L_u f =0 \nonumber
\end{equation}

This case is actually equivalent to a pair of first-order
equations
\begin{equation} \label{CS NLWE:f 2.1}
u_y=L, u_{z}=\frac{f-L_z}{L_u}.
\end{equation}

If we check directly the compatibility conditions they will coincide
with the determining equations of conditional invariance.

{\bf Case 2.1}. $K_u=0$, $K \ne 0$. The determining equations have
the form
\begin{equation}
-K L_{uu}=0, \nonumber
\end{equation}

\begin{equation}
L_{uy}-L_{uz}K+L_{uu}L - \frac{L_u K_y}{K}+\frac{K_y K_z}{K} -
K_{yz}  =0, \nonumber
\end{equation}

\begin{equation}
-f_y-Kf_z-Lf_u+L_{yz}+L_{uz}L+L_u f - \frac{K_y}{K}(L_z-f)-K_zf=0
\nonumber
\end{equation}

We have $K=k(y,z)$, $L=s(y,z)u+d(y,z)$. Using equivalence
transformations, we can put $d(y,z)=0$.

From the determining equations we get
\begin{equation}
k(y,z)=\frac{T_y}{T_z}, \label{CS NLWE:k}
\end{equation}
\begin{equation}
s(y,z)=\frac{T_{yz}}{T_z}, \label{CS NLWE:s}
\end{equation}
where $T=T(y,z)$ is an arbitrary function.

The operator of $Q$-conditional symmetry then will be
\[
Q=\partial_y + \frac{T_y}{T_z}\partial_z+\frac{T_{yz}}{T_z}u\partial_u.
\]

In this case the ansatz reducing equation (\ref{CS NLWE:yz-wave})
will have the form
\begin{equation} \label{CS NLWE:ansatz}
u=\sigma(y,z)\phi(\omega),
\end{equation}
\noindent
where $\omega=\omega(y,z)$ is a new variable,
\begin{gather}
T_y \omega_z+ T_z \omega_y=0, \nonumber \\
T_y \sigma_z+ T_z \sigma_y=\sigma T_{yz}. \nonumber
\end{gather}
The reduced equation will have the form:
\begin{equation} \label{CS NLWE:reduced}
\sigma_{yz}\phi + \phi'(\omega_y \sigma_z + \omega_z \sigma_y +
\sigma \omega_{yz}) +\phi''\sigma \omega_y \omega_z =f,
\end{equation}
\noindent
where $f$ satisfies the relevant conditions (\ref{CS NLWE:det
eq4}).

From these conditions we can find the form of the function
$f$ up to equivalence:
\begin{equation} \label{CS NLWE:f}
f=\frac{T_yT_z}{\sigma^3}\Phi(\omega,\frac{u}{\sigma})+\frac{\sigma_{yz}}{\sigma}u,
\end{equation}
\noindent
where $T(y,z)$ is the same arbitrary function entering expressions (\ref{CS NLWE:k}), (\ref{CS NLWE:s}).

At the first glance equation (\ref{CS NLWE:yz-wave}) may seem equivalent to some equation of the form
$f=T_yT_z\Phi(\omega,u)$ reducible with the ansatz $u=\phi(\omega)$. However, generally that is not the case.
The criterion for such reduction has the following form:
$$\sigma_y=k\sigma_z,$$
where $k$ is determined by (\ref{CS NLWE:k}).

Let us have a further look at the reduced equation (\ref{CS NLWE:reduced}).
From conditions on $\omega$ and $\sigma$ it is easy to check that
$$\omega_y \sigma_z + \omega_z \sigma_y + \sigma \omega_{yz}=0,$$
\noindent
so the reduced equation will have the form
$$\phi''\sigma \omega_y \omega_z = \frac{T_yT_z}{\sigma^3}\Phi(\omega,\phi).$$
Note that the reduced equations for this case will not include first-order derivatives.

As from conditions on $\omega$ and $\sigma$
$$\sigma^2=\frac{T_z}{\omega_z}=-\frac{T_y}{\omega_y},$$
\noindent
we come to the final form of the reduced equation

\begin{equation} \label{CS NLWE:red final}
\phi''=-\Phi(\omega,\phi).
\end{equation}

Equations of the form (\ref{CS NLWE:red final}) include many remarkable ODE, equations
for many special functions among them.

{\bf Case 2.2}. $K_u\ne 0$, then $K_{uu}K=K_u^2$,
$K=k(y,z)exp(l(y,z)u)$. We can put $k=1$ and prove from the
resulting determining conditions $l_y=l_z=0$, so we can put $l=1$.

Then we can found that $L=s(y,z)expu+d(y,z)$. It is possible to
reduce this case to $k=1$, and we get the following determining
equation for $f$ with arbitrary $s$ and $d$:

\begin{equation} \label{CS NLWE:f 3.1}
f=\frac{1}{3}(s_y+d_z),
\end{equation}
\noindent
so $f$ in this case depends only on $y$ and $z$, and the equation
$u_{yz}=f(y,z)$ is equivalent to the equation $u_{yz}=0$.

The conditions for $s$ and $d$ have the form
\begin{equation}
2s_{yz}-sd_z+2s_ys-d_{zz}=0,\\ -s_{yy}+2d_{yz}+s_yd-2d_zd=0.
\end{equation}

\section{Conclusions}
We have considered the equations
\begin{equation} \nonumber
u_{yz} = f(y,z,u)
\end{equation}
\noindent
with $f$ depending on $y,z,u$.

For such general class the only nontrivial case is Case 2.1, $K_u=0$, $K \ne
0$.

We have found that in this case the reduced equation has the form $\phi''=-\Phi(\omega,\phi)$,
including many remarkable equations.

We have found a general form of the equation (\ref{CS NLWE:yz-wave}) that
can be reduced to an ODE by means of an ansatz (\ref{CS NLWE:ansatz}) determined by the conditional
symmetry operator (\ref{CS NLWE:Q-operator a ne 0}) - $f$ has to be of the form (\ref{CS NLWE:f}).
However, for a general equation it may be not straightforward to determine
whether $f$ can be reduced to such form.

The cases $f=f(u)$, $f=r(y,z)u$ require special consideration, and have more
inequivalent cases.

Further research may also include study of the general conditional symmetry
of the nonlinear wave equation in higher dimensions, as well as description of equivalence classes
of conditional symmetries.

}
\end{document}